\documentclass[sigconf]{acmart}

\acmConference[MSR 2026]{MSR '26: Proceedings of the 23rd International Conference on Mining Software Repositories}{April 2026}{Rio de Janeiro, Brazil}
\setcopyright{rightsretained}
\copyrightyear{2026}
\acmYear{2026}
\acmDOI{XXXXXXX.XXXXXXX}

\usepackage{graphicx} 
\usepackage{subcaption}

\usepackage{hyperref}
\usepackage{listings}
\usepackage{float}
\usepackage{fancyvrb}
\usepackage{tabularx}
\usepackage{wrapfig}
\usepackage{tikz-imagelabels}
\usepackage{natbib}
\setcitestyle{square,citesep={,}}
\usepackage{multirow}
\hypersetup{hidelinks,linkcolor = black} 
\usepackage{tabularx}
\usepackage{booktabs}

\usepackage{numprint}
\npthousandsep{,}
\npdecimalsign{.}
\nprounddigits{1}
\usepackage[export]{adjustbox}
\usepackage[most]{tcolorbox}
\usepackage{xcolor}
\newtcolorbox{issuequote}{
    enhanced,
    breakable,
    colback=white,
    colframe=black!60,
    boxrule=0pt,
    borderline west={1.4pt}{0pt}{black!65},
    sharp corners,
    left=1em,
    right=0.5em,
    top=0.3em,
    bottom=0.3em,
    before skip=8pt,
    after skip=10pt,
    fontupper=\itshape
}

\newtcolorbox{mainfinding}[1]{
    enhanced,
    breakable,
    colback=white,
    colframe=black!65,
    boxrule=0.5pt,
    sharp corners,
    title=\textbf{#1},
    fonttitle=\small,
    coltitle=black,
    left=1em,
    right=1em,
    top=0.8em,
    bottom=0.8em,
    before skip=8pt,
    after skip=8pt
}

\begin{document}

\title{Characterizing Visual Accessibility Issues in AI Developer Tools: An Empirical Study}
\author{Sabrina Haque}
\affiliation{%
  \department{Department of Computer Science and Engineering}
  \institution{University of Texas at Arlington}
  \city{Arlington}
  \state{Texas}
  \country{USA}}
\email{sxh3912@mavs.uta.edu}

\author{Christoph Csallner}
\affiliation{%
  \department{Department of Computer Science and Engineering}
  \institution{University of Texas at Arlington}
  \city{Arlington}
  \state{Texas}
  \country{USA}}
\email{csallner@uta.edu}

\begin{abstract}
AI-assisted developer tools increasingly mediate programming through chat panels, terminal agents, generated diffs, and streaming status output. These interaction surfaces may create visual accessibility barriers for blind, low-vision, and color-vision-deficient developers, yet little is known about how such barriers are reported in public tool ecosystems. We analyze issues and forum discussions from five AI developer tool ecosystems: GitHub Copilot in VS Code, Cursor, Claude Code, OpenAI Codex, and OpenCode. From 2,652 keyword-retrieved candidates, a three-model ensemble identified 600 unanimously positive visual accessibility reports. A stratified manual
sanity check supported this conservative selection. Topic modeling and qualitative analysis identified three recurring categories: screen-reader and assistive-technology barriers; visual presentation, contrast, and differentiation problems; and readability, scaling, and control limitations in AI-specific interfaces. The relative prominence of these concerns varied across ecosystems and reflected differences in editor, terminal, chat, diff, and agent interaction surfaces. An exploratory metadata analysis further identified differences in
reporter activity and, across the GitHub-based ecosystems, maintainer participation and closure processes. These findings show that the accessibility record of AI developer tools is shaped by both their interaction design and the reporting and maintenance practices of their surrounding ecosystems.

\end{abstract}

\maketitle

\section{Introduction}

The rapid integration of generative AI has led to a shift in the software development landscape. Developers now use tools such as GitHub Copilot~\cite{copilot} and Cursor~\cite{cursor} to write code, explain programs, debug errors, and modify existing projects~\cite{liang2024large}. These tools provide chat panels, inline assistance, terminal command execution, and act as semi-autonomous agents that work directly inside the editor or terminal~\cite{vscode_agentmode}. Rather than writing and executing code directly, developers increasingly guide AI through natural language prompts and iteratively refine AI-generated outputs~\cite{barke2023grounded,peng2023impact}. This
shift creates new opportunities for software development, but it also
raises an important question: whether these emerging AI-assisted tools are usable and accessible for all developers.

Visual accessibility concerns whether interfaces can be perceived, understood, and used by people with visual impairments, including
blindness, low vision, and color-vision deficiency. Programming environments are already difficult to use with screen readers~\cite{mountapmbeme2022addressing}. Visually impaired developers often face barriers in code navigation, debugging, and collaborative software work~\cite{albusays2017interviews}. Recent studies show that generative AI can help by supporting debugging, explaining code, describing visual materials, and reducing dependence on sighted colleagues~\cite{cha2025game}. However, the same tools can also introduce new risks, including inaccessible interfaces, auditory overload, and difficulty interpreting generated outputs~\cite{flores2025impact, tang2025everyday}. These challenges become especially important because AI coding tools introduce dynamic content, such as real-time code generation and streaming responses, along with complex interactions that traditional accessibility practices may not fully address~\cite{chen2026programmers}. If the interfaces around AI assistance are not accessible, a tool meant to speed developers up can instead slow them down or exclude them from emerging software-development workflows. As AI coding tools become basic infrastructure for software work, their accessibility must be treated as a core requirement rather than a secondary usability concern.

Accessibility problems in AI tools are not hypothetical~\cite{adnin2024look}. A preliminary screen reader evaluation of ChatGPT, Copilot, Gemini, and Perplexity found usability difficulties across tools, especially in navigation structure, labeling, feedback mechanisms, and prompt handling~\cite{leporini2025preliminary}. Similar tension exists in coding and developer tools. Recent studies of screen reader users and AI coding assistants found that these tools can empower users, but users still struggle to communicate intent, review generated output, switch between views, and retain control~\cite{chen2026programmers,flores2025impact}.

These studies provide deep insight into user experience through
interviews, controlled tasks, and expert evaluation. However, they do not show how visual accessibility problems surface in the public maintenance channels where AI developer tools evolve. Issue trackers and discussion forums capture bug reports, feature requests, accessibility testing findings, and user discussions as problems are encountered in practice. Mining these artifacts has already proven useful for studying accessibility gaps in software projects and developer communities~\cite{bi2021first,alghamdi2025understanding}. However, existing mining work has largely focused on general software, mobile apps, or broad accessibility discussions. To our knowledge, no prior mining study has focused on visual accessibility reports in AI developer tools. This gap matters because AI developer tools introduce interaction surfaces that differ from traditional editor, web, or mobile interfaces, including chat panels, streaming responses, generated diffs, terminal transcripts, and agent status updates.

In this work, we study visual accessibility reports from five popular AI developer-tool ecosystems. We selected these ecosystems to represent different primary interaction environments and because each provides a public channel through which users report problems. \textbf{GitHub Copilot in VS Code}~\cite{copilot} represents editor-integrated AI assistance. GitHub Copilot is the most widely adopted AI coding assistant, with over 20 million cumulative users~\cite{copilot_users}, and is deeply integrated into VS Code, the most widely used code editor~\cite{vscode_survey}. \textbf{Cursor} represents an AI-native code editor and has grown to over one million daily active users~\cite{cursor_users}. \textbf{Claude Code}~\cite{claude_code} and \textbf{OpenCode}~\cite{opencode} prominently support terminal-oriented agent workflows, whereas \textbf{OpenAI Codex} spans command-line, desktop, and editor interfaces~\cite{codex}. These descriptions refer to the primary interaction environments represented by these tools but several tools support multiple interfaces. OpenCode is an open-source coding agent with over 184k GitHub stars~\cite{opencode} and 23k forks, and its public development process allows us to observe how accessibility concerns are raised and addressed in a community-maintained project. The AI developer-tool landscape is large and evolving quickly; together, these ecosystems provide a diverse view of accessibility concerns across prominent editor-integrated, AI-native, terminal-oriented, and multi-interface development environments.

We mined bug reports and forum discussions from each tool's own public maintenance channels. For Claude Code, OpenAI Codex, and OpenCode, we collected GitHub issues.\footnote{\url{https://github.com/anthropics/claude-code}}\footnote{\url{https://github.com/openai/codex}}\footnote{url{https://github.com/anomalyco/opencode}}
GitHub Copilot does not maintain a separate public issue tracker, so we queried the VS Code repository\footnote{\url{https://github.com/microsoft/vscode}} for issues mentioning Copilot created after its public launch in June 2021. Cursor does not use a public GitHub issue tracker for user reports, so we collected threads from the official Cursor community forum.\footnote{\url{https://forum.cursor.com/}} We included all reports created on or before June 15, 2026.  
We used each tool's public issue tracker or official forum because these channels record problems reported directly to maintainers. Using a mixed-method analysis, we characterize the visual accessibility concerns surfaced across these tools' public ecosystems. We mainly ask two research questions:

\begin{itemize}
    \item \textbf{RQ1}: What types of visual accessibility issues are reported in the public channels of AI-assisted developer tools?
    \item \textbf{RQ2}: How do the nature and response patterns of reported visual accessibility barriers vary across AI developer tool ecosystems?
\end{itemize}

\section{Methodology}

\begin{figure}[h!t]
    \centering
    \includegraphics[width=\linewidth]{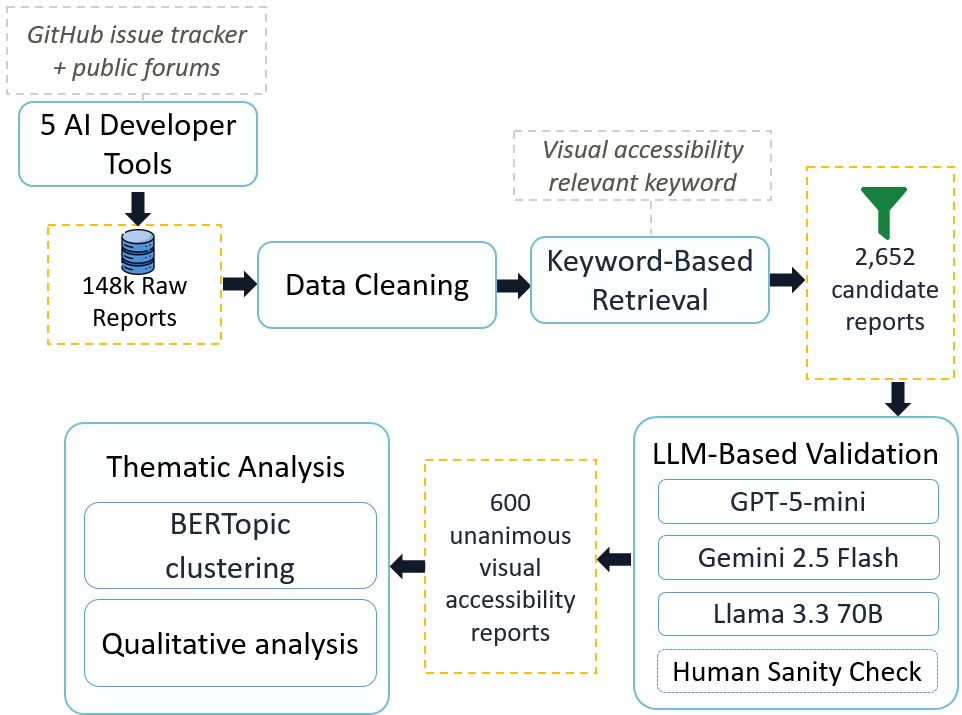}
    \caption{Overview of the study methodology}
    \label{fig:workflow}
\end{figure}

To answer our research questions, we conducted a mixed-methods empirical study of visual accessibility reports from public AI developer-tool ecosystems. Figure~\ref{fig:workflow} summarizes the study pipeline. We collected reports from the five ecosystems (GitHub Copilot in VS Code~\cite{copilot}, Cursor~\cite{cursor}, Claude Code~\cite{claude_code}, OpenAI Codex~\cite{codex}, and OpenCode~\cite{opencode}) as introduced earlier, which represent editor-integrated, AI-native, terminal-oriented, and multi-interface development environments. We then used BERTopic~\cite{grootendorst2022bertopic} and qualitative analysis to identify recurring accessibility barriers and compare their distribution across tool ecosystems. For RQ2, we additionally used issue metadata collected with the reports to examine reporter activity and repository-level response patterns.

\subsection{Data Collection and Preprocessing}

We collected public reports created on or before June~15, 2026. For GitHub Copilot in VS Code, Claude Code, OpenAI Codex, and OpenCode, we collected publicly available GitHub issues through the GitHub REST API~\cite{github_api}. We retrieved issues regardless of state and retained the title, body,
labels, state, reporter, creation date, comments, comment-author associations, closure time, closure actor, and closure reason when available. 

GitHub Copilot does not maintain a separate public issue tracker. We therefore queried the VS Code repository for issues containing the term \texttt{copilot} that were created after Copilot's public launch in June 2021. After accessibility candidate retrieval, we manually reviewed the retained reports to exclude issues that mentioned Copilot but concerned general VS Code behavior rather than Copilot functionality. Cursor likewise does not use a public GitHub issue tracker for user reports. We collected public threads returned by searches of the Cursor community forum~\cite{cursor_forum}, retaining each thread's title, body, tags, reply count, and creation date. Forum search served only to identify potential reports; all retrieved threads underwent the same cleaning and validation procedure as the GitHub issues.

We implemented the entire data-processing pipeline in Python. Before candidate retrieval, we removed duplicate issue reports and reports with no body text. Our analysis was limited to issue reports and excluded pull requests. We removed non-English reports using the \texttt{langdetect}~\cite{langdetect} library. We then cleaned the title and body text using regular-expression-based preprocessing. This process removed URLs, email addresses, user mentions, HTML tags, code blocks, inline code, Markdown formatting, quoted text, embedded links and images, and
configuration tables or boilerplate commonly included in issue templates. We normalized whitespace and converted the remaining text to lowercase. These steps ensured that keyword matching in the subsequent step operated on the natural language content of each issue rather than on code, metadata, or formatting artifacts~\cite{sanei2024characterizing}. 

\subsection{Keyword-Based Candidate Retrieval}
\label{sec:keywords}

The full collection across five tools contains a large volume of reports on topics unrelated to accessibility, including functional bugs, feature requests, performance concerns, and documentation questions. Rather than validating every report with a language model, which would be costly and wasteful at this scale, we used keyword-based retrieval as a first-pass filter. This step discarded off-topic content cheaply and reproducibly, leaving a smaller candidate pool of potentially visual-accessibility-related reports. Keyword-based filtering is a well-established first step in mining studies of accessibility reports, which commonly use lexical retrieval to construct candidate sets before finer-grained classification~\cite{oliveira2023analyzing}.

We developed the keyword set by drawing on prior empirical studies of accessibility discussions in developer communities and issue trackers~\cite{huq2023a11ydev, bi2021first, alghamdi2025understanding}. Rather than using a flat list, we organized the keywords into semantic groups:

\textit{\textbf{General accessibility:}} \texttt{accessibilit*}, \texttt{a11y}, 
\texttt{no accessible}, \texttt{not accessible}, \texttt{inaccessible}, \texttt{disabilit*}

\textit{\textbf{Visual impairment:}} \texttt{blind}, \texttt{low vision}, \texttt{visually impaired}, \texttt{vision impaired}, \texttt{vision impairment}, \texttt{visual impairment} 

\textit{\textbf{Assistive technology or rules:}} \texttt{screen reader}, \texttt{screen-reader}, \texttt{screenreader}, \texttt{nvda}, \texttt{jaws}, \texttt{voiceover}, \texttt{voice over}, \texttt{voice access},  \texttt{assistive technology}, \texttt{braille}, \texttt{wcag}, \texttt{section 508} 

\textit{\textbf{Labeling and alt text:}} \texttt{aria}, \texttt{aria-label}, 
\texttt{accessible name}, \texttt{accessible label}, \texttt{alt text}, 
\texttt{alternative text}, \texttt{alt-text}, \texttt{content description} 

\textit{\textbf{Visual presentation:}} \texttt{contrast}, \texttt{color contrast}, 
\texttt{color blind}, \texttt{colorblind}, \texttt{font size}

We used wildcard matching for the stemmed terms \texttt{accessibilit*} and \texttt{disabilit*} to capture morphological variants. Matching was case-insensitive and applied to the cleaned title and body of each report.
We performed keyword retrieval only after cleaning the reports. Raw issue and forum content often contains code blocks, issue templates, copied logs, etc. which may include accessibility-related terms, unrelated to the reporter's actual concern. For example, an issue template may contain a generic accessibility field, while a code snippet may include an \texttt{aria-label} attribute unrelated to the reported problem. Applying keyword matching after cleaning is aimed to reduce this noise.
Cleaning alone could not eliminate all false positives. Terms such as \textit{contrast}, \textit{blind}, and \textit{accessible} may still be used in other contexts. We therefore treated the retrieved reports as candidates for further validation. \textbf{This process produced 2,652 candidate reports across the five tool ecosystems} (Table~\ref{tab:llm-agreement}, column ``N'').

\subsection{LLM-Based Validation}
\label{sec:classification}

The keyword retrieval step identified reports that potentially concerned visual accessibility, but it could not determine whether accessibility was the reporter’s actual concern. Since manual annotation was not feasible at this point, we opted for an LLM-based approach. Recent work suggests that LLMs can support qualitative software engineering research, but should be assessed through agreement and validation procedures~\cite{ahmed2025can,bano2024large}. We therefore used an ensemble of three large language models to validate the retrieved candidates and annotate whether each report described a visual or screen-reader accessibility barrier.  

Each model received the same structured prompt containing the cleaned issue title and body. The prompt defined visual accessibility issues using explicit inclusion and exclusion criteria grounded in prior accessibility research~\cite{mountapmbeme2022addressing,alghamdi2025understanding}. Listing~\ref{lst:prompt} presents the complete prompt.

We selected GPT-5-mini~\cite{gpt5}, Gemini 2.5 Flash~\cite{gemini_flash}, and Llama 3.3 70B~\cite{llama} to include models from different providers and model families. Using multiple models reduced risk of model-specific biases and enabled us to assess the consistency of the resulting annotations.

Each model returned a structured JSON response containing a three-way label (\texttt{yes}, \texttt{no}, or \texttt{uncertain}), a binary accessibility flag and a list of key terms. Following prior work, we also included a confidence score from 0 to 100 and a brief rationale to make the model outputs easier to inspect and interpret~\cite{dev2025beyond}. The rationale helped record the basis for each classification, while the confidence score captured the model's self-reported certainty. 

\begin{figure}[h!t]
\begin{lstlisting}[
  caption={Prompt template used for LLM-based classification.},
  label={lst:prompt},
  basicstyle=\ttfamily\scriptsize,
  breaklines=true,
  breakatwhitespace=false,
  frame=single,
  numbers=none,
  columns=flexible
]
You are a software developer and accessibility researcher analyzing GitHub issues from AI coding tools. Your task is to determine whether the issue is specifically about vision-related accessibility.

Definition: An issue is vision-accessibility-related if it concerns barriers affecting blind, low-vision, color-blind, or visually impaired users. This includes:
- Screen reader or assistive technology problems (e.g., NVDA, JAWS, VoiceOver, Narrator)
- Missing or incorrect accessible names, labels, ARIA, roles, semantic structure, or announcements
- Color contrast, high-contrast mode, dark/light theme readability, or color-only information affecting accessibility
- Font size, zoom, layout readability, or visual clarity when it concerns accessibility barriers for low-vision users
- Focus visibility or focus behavior when it affects accessibility, screen reader use, or low-vision users
- Accessibility of AI-generated output (chat responses, streaming text, terminal output, progress indicators, diffs, status messages) when it affects screen reader 
  or low-vision users

Exclude:
- General keyboard shortcut bugs with no visual or screen-reader accessibility impact
- General UI bugs, layout preferences, or theme requests that do not specifically concern visual accessibility
- General terminal bugs unless the issue concerns screen reader access, contrast, ANSI/color output, spinners, or visual accessibility
- Feature requests unrelated to vision or screen-reader accessibility
- Issues where accessibility is mentioned only in unrelated boilerplate or templates

Issue Title: {title}
Issue Body:  {body}

Respond with ONLY a valid JSON object. Do not include markdown or extra text. Use this schema:
{
  "label": "yes" | "no" | "uncertain",
  "is_visual_or_screen_reader_accessibility": true | false,
  "confidence": 0-100,
  "key_terms": ["term1", "term2"],
  "reasoning": "Brief explanation in one sentence."
}

Rules:
- Use "yes" only when the issue clearly concerns visual or screen-reader accessibility.
- Use "no" when the issue is clearly unrelated.
- Use "uncertain" when the issue may be accessibility-related but lacks enough evidence.
- Set is_visual_or_screen_reader_accessibility to true only for "yes"; otherwise false.
- Keep reasoning brief.
\end{lstlisting}
\end{figure}

Across the 2,652 keyword-retrieved candidates, all three models agreed on the binary label for 2,239 reports (84.4\%). The ensemble produced 600 unanimously positive reports and 1,639 unanimously negative reports; the remaining 413 reports received mixed labels. To assess how consistently the models applied the classification criteria, we computed inter-model agreement using Cohen's kappa for each pair of models~\cite{mchugh2012interrater}. The pooled pairwise Cohen’s $\kappa$ values ranged from 0.75 to 0.76, which falls in the substantial-agreement range, indicating that the three models applied the accessibility criteria consistently despite coming from different model families. Table~\ref{tab:llm-agreement} summarizes the agreement and retained reports by tool ecosystem.

\begin{table*}[h!t]
\centering
\caption{Inter-model agreement and high-confidence visual accessibility reports by tool ecosystem.
N = issues per repository after keyword-based filtering; GPT = GPT-5-mini,
Gemini = Gemini 2.5 Flash, Llama = Llama 3.3 70B. ``Uncertain'' labels are
treated as ``No''. Agr.\ = percentage of issues for which all three models produced the same
binary label; Unan.\ = issues unanimously labeled yes or no; $\kappa$ = Cohen's Kappa;
Combined $\kappa$ values are pooled across repositories.}
\label{tab:llm-agreement}

\setlength{\tabcolsep}{4pt}
\begin{adjustbox}{max width=\textwidth}
\begin{tabular}{lrrrrrrrrrr}
\toprule
& & \multicolumn{3}{c}{\textbf{``Yes''}} & &
\multicolumn{2}{c}{\textbf{Unan.}} &
\multicolumn{3}{c}{\textbf{Cohen's $\kappa$}} \\
\cmidrule(lr){3-5} \cmidrule(lr){7-8} \cmidrule(lr){9-11}
\textbf{Tool} & \textbf{N} &
\textbf{GPT} & \textbf{Gemini} & \textbf{Llama} &
\textbf{Agr.\ (\%)} &
\textbf{Yes} & \textbf{No} &
\textbf{GPT--Gem.} & \textbf{GPT--Llama} & \textbf{Gem.--Llama} \\
\midrule
VS Code Copilot & 309  & 142 & 180 & 177 & 72.5 & 122 & 102  & 0.76 & 0.52 & 0.62 \\
Claude Code     & 1430 & 308 & 412 & 319 & 87.7 & 263 & 991  & 0.76 & 0.79 & 0.78 \\
OpenAI Codex    & 318  & 92  & 112 & 85  & 84.0 & 71  & 196  & 0.77 & 0.76 & 0.72 \\
OpenCode        & 215  & 67  & 84  & 67  & 85.6 & 57  & 127  & 0.77 & 0.80 & 0.79 \\
Cursor Forum    & 380  & 102 & 139 & 116 & 81.6 & 87  & 223  & 0.71 & 0.76 & 0.69 \\
\midrule
\textbf{Combined} & 2652 & 711 & 927 & 764 & 84.4 & 600 & 1639 & 0.76 & 0.75 & 0.75 \\
\bottomrule
\end{tabular}
\end{adjustbox}
\end{table*}

We retained only the 600 unanimously positive reports as a conservative high-confidence set for thematic analysis. Some relevant reports may have been excluded when one model was uncertain or disagreed, but the retained set provides a stronger basis for analyzing recurring issue types.

Manual validation of all 2652 issues across 5 repositories was not feasible, so we conducted a stratified manual sanity check to assess whether the ensemble’s labels aligned with human judgment. We sampled 100 reports across the five tool ecosystems: 50 unanimously positive reports, 25 unanimously negative reports, and 25 reports for which the models disagreed. The sample included 20 reports from each ecosystem and was balanced across the three agreement conditions. Each report was reviewed manually without access to the model labels and assigned one of three labels: yes, no, or uncertain. 

Among the 50 unanimously positive reports, 47 were judged visual-accessibility related, two were judged uncertain, and one was judged unrelated. All 25 unanimously negative reports were judged unrelated. The disagreement sample was mixed: 11 reports were judged related, 13 unrelated, and one uncertain. These results support the use of unanimous positive agreement as a conservative selection criterion. They also show that the reports on which the models disagreed were genuinely ambiguous rather than straightforward cases that the ensemble failed to classify consistently.

\subsection{Thematic Clustering and Qualitative Analysis}
\label{sec:clustering}

We used BERTopic~\cite{grootendorst2022bertopic} to organize the high-confidence reports into recurring thematic groups before conducting qualitative analysis. BERTopic was suitable for this task because the reports were relatively short, varied in wording across tool ecosystems, and often used to find themes in similar online discussions~\cite{opu2025understanding}. The method combines sentence embeddings with density-based clustering and class-based TF-IDF representations, allowing semantically related reports to be grouped while retaining interpretable topic terms~\cite{mcinnes2017hdbscan,reimers2019sentence}.

We applied BERTopic to the cleaned title and body text of the 600 high-confidence reports. We used \texttt{all-mpnet-base-v2} to generate embeddings and set the minimum topic size to 15. This threshold reduced the likelihood of treating small, isolated groups of reports as meaningful themes. We treated the clustering output as an exploratory structure rather than a final taxonomy. The purpose of clustering was to identify candidate thematic groupings that could be examined and interpreted through manual analysis.

BERTopic produced three non-outlier clusters containing 262, 215, and 90 reports, respectively. The remaining 33 reports were assigned to BERTopic’s outlier topic. To interpret and characterize each cluster, we manually reviewed 50 randomly selected reports from each cluster, resulting in a qualitative sample of 150 reports. Sampling was stratified by tool ecosystem. For each sampled report, we recorded the primary accessibility barrier, the interaction surface where it occurred, and a finer-grained subtheme via inductive coding. Table~\ref{tab:topics} presents representative top terms for each
cluster and the labels we assigned after reviewing sampled reports.

\begin{table}[t]
\caption{BERTopic clusters and qualitative interpretations
(\(N=600\) unanimously classified accessibility reports). Top terms
are shown after preprocessing and are used only to summarize the
automatically derived clusters.}
\label{tab:topics}
\centering
\small
\begin{tabular}{p{0.11\linewidth} p{0.10\linewidth} p{0.34\linewidth} p{0.33\linewidth}}
\toprule
\textbf{Cluster} & \textbf{Reports} & \textbf{Representative top terms} & \textbf{Qualitative interpretation} \\
\midrule
0 & 262 & screen, accessibility, reader, screen reader, chat, version &
Screen reader and assistive technology \\
1 & 215 & theme, color, dark, terminal, contrast, background &
Visual presentation, contrast, and differentiation \\
2 & 90 & font, size, font size, chat, settings, ui &
Readability, scaling, and interface control \\
Outliers & 33 & --- & Reports not assigned to a stable topic \\
\bottomrule
\end{tabular}
\end{table}

\section{Findings}
\label{sec:findings}

We use the three clusters and our qualitative analysis of sampled reports to answer \textbf{RQ1} by characterizing the visual accessibility barriers reported across AI developer tools. We then answer \textbf{RQ2} by comparing how these barriers are distributed across the five tool ecosystems.

\subsection{RQ1: Types of Reported Visual Accessibility Barriers}

Our qualitative analysis of the three BERTopic-derived clusters
identified three recurring categories of visual accessibility barriers: (1) screen reader and assistive technology barriers, (2) visual presentation, contrast, and differentiation problems, and (3) readability, scaling, and control limitations in AI-specific interfaces. The following subsections explain how these barriers manifested in the sampled reports and the interaction surfaces where they occurred.

\subsubsection{Screen Reader and Assistive Technology}

The largest cluster (262 reports) concerns screen-reader and assistive-technology access. In manually analyzed reports, most entries involved screen readers, including NVDA, JAWS, VoiceOver, Narrator, and ORCA. A smaller set raised voice access, text-to-speech, keyboard-focus, or motion-related concerns. Users were concerned with accessing different points in an AI-assisted workflow, such as configuring a request, working with generated output, and monitoring an agent while it runs.

A recurring concern involved AI-specific controls. Reports described prompt fields whose commands or context options were not announced, model selectors whose current or expanded state was unclear, and
dialogs whose keyboard focus moved away before users could reach permission or action controls. For example, one Copilot accessibility report noted that NVDA did not announce the inline-chat placeholder, which described options for adding context, accessing extensions, and invoking
commands. Other reports identified inconsistent shortcuts for adding context across chat surfaces and action buttons that appeared only on hover. The latter also created a barrier for voice access because a button that is not discoverable cannot be invoked. These reports show that AI controls must clearly expose available commands, state, and interaction options so that developers using assistive technologies can access the tool's full functionality.

Another concern was the gap between reading AI output and acting on it. Reports described inline suggestions that were visible but not exposed to NVDA, chat responses without headings or landmarks, and code blocks without clear labels or group information. These gaps made it difficult to find the latest response, distinguish suggestions, or move through long conversations. One NVDA user reported:

\begin{quote}
    \textit{``Multiple accessibility problems when using a screen reader (NVDA). The main one is that is is impossible to navigate to the last response. reading up from the input field using NVDA browse mode jumps far back in the chat responses. reading down from there does show later entries in the chat, but there's no easy way to get to the last response without going line by line.''}
    \footnote{\url{https://github.com/openai/codex/issues/26491}}
\end{quote}

Generated output also needed to remain actionable, not merely readable. One report described an Accessible View in which a user could reach a generated code block but could not complete the next step of inserting it into the editor:

\begin{quote}
  \textit{``This is not currently supported and users have to close the Accessible View and manually relocate to the codeblock.''}
  \footnote{\url{https://github.com/microsoft/vscode/issues/199119}}
\end{quote}

Thus, access to generated text alone was insufficient. Developers also needed to reach the actions associated with that output from the accessible representation.

Terminal-based agents raised another set of concerns because their status changes while the model is working. Reports described spinners, streaming text, and repeated redraws that produced confusing speech, skipped lines, interrupted reading, or moved the terminal viewport. One visually impaired user explained:

\begin{quote}
\textit{``I am visually impaired and operate the terminal app using Apple’s VoiceOver screen reader...At present there doesn’t seem to be a way to know when it has finished generating a response. As a result, I periodically copy the terminal output into another text editor and have it read aloud to check for any differences.''}\footnote{\url{https://github.com/openai/codex/issues/150}}
\end{quote}

Other sampled reports described animated status labels that generated unhelpful speech output, redraws that moved the terminal viewport away from earlier content, and streaming behavior that caused screen readers to freeze or skip lines. A smaller set of reports requested read-aloud responses and built-in voice access.

\subsubsection{Visual Presentation, Contrast, and Differentiation}

The second cluster (215 reports) concerns the visual presentation of AI-specific interfaces. In the manual sample, reports described low-contrast text, weak visual separation between different types of content, and limited control over colors and themes. These problems appeared in chat panels, terminal transcripts, agent lists,
tool output, and generated diff views.

A recurring concern involved text or highlights that became difficult or impossible to read under particular terminal or editor themes. Reports described syntax-highlighted code, tool results, selected items, links, and approval dialogs whose foreground and background colors had insufficient contrast. In some cases, text was present but could only be read after selecting it:

\begin{quote}
    \textit{``Kotlin annotations in code blocks are rendered in a color that is identical or nearly identical to the background on dark terminal themes. The text is present (visible when selecting/highlighting) but completely unreadable without intervention.''}
    \footnote{\url{https://github.com/anthropics/claude-code/issues/24807}}
\end{quote}

These cases were not limited to a single appearance mode. They included light themes in which text inherited a light foreground, dark terminal themes in which syntax-highlighted code blended into the background, and high-contrast editor themes in which inline AI content remained invisible.

A related set of reports concerned the lack of control over these visual encodings. AI developer tools often introduce their own colors for prompts, tool calls, agent names, status messages, and diffs. When these colors are hardcoded or only partly inherit the host theme, users cannot adapt the interface to their visual needs. Sampled reports requested configurable foreground and background colors, theme-aware rendering, and color schemes that respect terminal palettes. Many reports also requested theme customization or compatibility with existing terminal and editor settings. 
Diff rendering was a particularly common example. Muted colors might blend with syntax highlighting, making added and removed lines harder to inspect while hardcoded bright colors could clash with dark terminals and cause visual fatigue during long review sessions. Some reports requested a minimal diff mode that used non-background indicators, such as line markers or line numbers, rather than relying on large blocks of fixed color.
One feature request from a colorblind developer stated:

\begin{quote}
    \textit{``I would like to suggest / request adding Colorblind Theme / Mode to the Web / Desktop app.
    My main pain as a Colorblind is distinguishing between the Add / Removed lines in the Diff View.''}
    \footnote{\url{https://github.com/anomalyco/opencode/issues/14755}}
\end{quote}

Another pattern involved weak visual distinction between the parts of an AI interaction. In long chat or terminal sessions, user prompts, assistant responses, tool output, and agent status lines could use nearly identical styling. This made it difficult to find a previous instruction, identify the source of a message, or follow the flow of
an extended interaction:

\begin{quote}
\textit{``User prompts and assistant responses are currently rendered with very little visual distinction in the main conversation flow.
At the moment, this makes long sessions harder to scan, especially when looking back through a discussion to find a previous user instruction.''}
\footnote{\url{https://github.com/openai/codex/issues/17879}}
\end{quote}

These reports show that visual hierarchy is especially important in AI-assisted tools, where developers must repeatedly separate their own instructions from generated responses, tool activity, and proposed changes.

\subsubsection{Readability, Scaling, and Control}

The third cluster (90 reports) concerns the readability and customization of AI-specific interfaces. In the manual sample, most reports involved font size, zoom, typography, and small interface elements. Others concerned dense layouts and weak hierarchy in long AI-generated responses. These issues appeared in chat panels, plan views, terminal output, agent windows, and generated code or documentation.

The most common concern was that AI panels did not provide independent control over text size or typography\footnote{\url{https://forum.cursor.com/t/font-size-in-chat/56747}}. Reports described chat text that was smaller than the editor text, fixed font families that were hard to read, and AI panels that ignored system or editor font settings. In many cases, the only workaround was to zoom the entire application. This changed the rest of the workspace even when developers only needed larger text in the chat or agent panel. One feature request described the impact on a developer with astigmatism:

\begin{quote}

    \textit{``The default font it responds with is too small and blurry for my astigmatism. It needs to be better for vision because my vision isn't that bad. I know many developers with far worse vision than me, and if I'm struggling to read the output  then it's nearly inaccessible to them.''}
    \footnote{\url{https://github.com/anthropics/claude-code/issues/44248}}
\end{quote}

Several reports showed that enlarging text could create a second problem. Global zoom could hide tooltips or controls, crop overlays, or reduce the usable space for code and conversation. Large fonts also made fixed headers, decorative elements, and multi-line status areas consume much of a terminal or chat viewport. One visually impaired user explained:

\begin{quote}
    \textit{``I use very large fonts and it compresses down the scroll area a lot when there is a lot of text at the top.''}
    \footnote{\url{https://github.com/anomalyco/opencode/issues/2750}}
\end{quote}

The report requested a compact layout that could hide decorative elements and preserve space for the conversation. These cases show that supporting larger text requires responsive layouts, not only a
font-size setting.

Another pattern concerned the readability of long AI-generated responses. Chat interfaces often present plans, summaries, and step-by-step instructions using Markdown headings and lists. However, reports described headings that were visually indistinguishable from body text, fixed line height and paragraph spacing that made responses look like dense walls of text, and truncated agent names that made it hard to identify the active agent. One report stated:

\begin{quote}
\textit{``All markdown headings (\texttt{h1}--\texttt{h6}) rendered
inside chat messages are visually identical to body text.''}
\footnote{\url{https://github.com/anomalyco/opencode/issues/30542}}
\end{quote}

When response hierarchy is not visible, developers must scan long generated text line by line to locate plans, decisions, or next steps. This is particularly problematic in AI-assisted workflows, where generated responses often serve as the main place for explaining actions, presenting alternatives, and summarizing work.

\begin{figure}
    \centering
    \includegraphics[width=\linewidth]{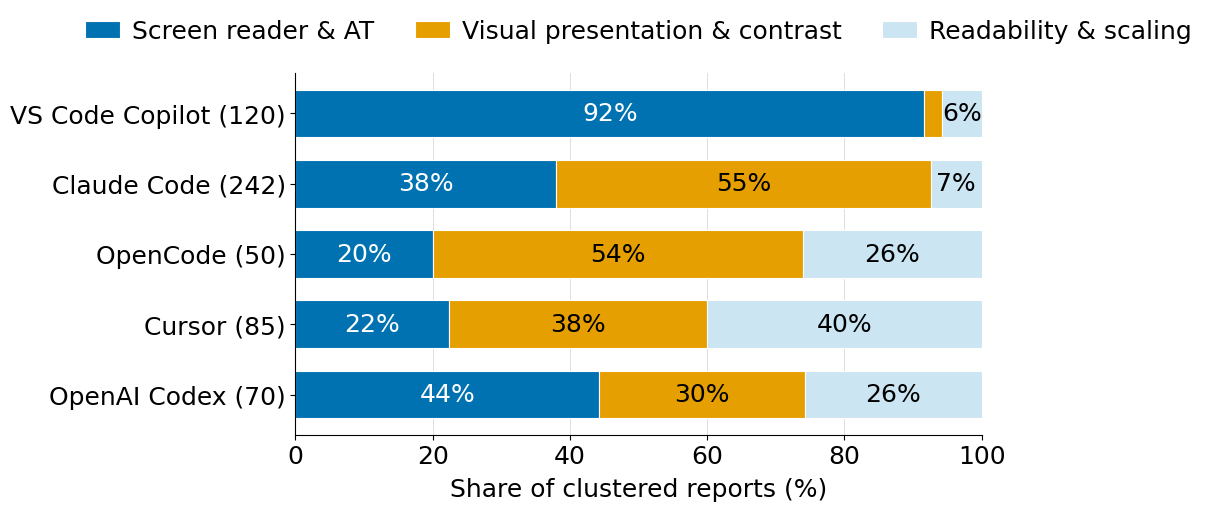}
    \caption{Distribution of visual accessibility issue types across AI developer tools; shows the distribution of the 567 reports assigned to one of the three BERTopic clusters. Thirty-three reports assigned to the outlier topic are excluded. Percentages below 3\% are not labeled.}
    \label{fig:clusters}
\end{figure}

\subsection{RQ2: Nature and Response Patterns of Reported Visual Accessibility Barriers Across Tools}

RQ2 combines three sources of evidence. Figure~\ref{fig:clusters} uses the 567 reports assigned to the three substantive BERTopic clusters to compare the distribution of barrier categories across
tools. The remaining 33 reports were excluded from this comparison because BERTopic did not assign them to a coherent cluster. However, they were included in the metadata analyses because their outlier status reflects topic assignment rather than report validity. These distributions indicate which barriers were most visible in each tool's public reports. Our interpretation of the variation
across tools draws on the 150 manually reviewed reports used in RQ1.

We also use issue metadata from all 513 reports in the four GitHub-based ecosystems to examine comments, maintainer participation, issue status, and closure patterns. Cursor is excluded from the response analysis because forum threads do not provide issue states, closure records, or author-association metadata directly comparable to GitHub. These measures describe visible activity in the repositories and do not indicate whether a reported barrier was fixed. We classified accounts as bots when the login included a bot marker, such as a ``[bot]'' suffix. We treated comments from accounts with the \texttt{OWNER}, \texttt{MEMBER}, or \texttt{COLLABORATOR} association as maintainer participation.

\subsubsection{Nature of Reported Barriers}
VS Code Copilot reports were dominated by screen-reader and assistive-technology barriers (92\%). These reports were concentrated in editor-integrated AI surfaces, including Copilot Chat, Quick Chat, inline chat, suggestions, and Accessible View. Users described difficulty determining the state of model and context controls, locating a newly generated response, and moving from an accessible representation of generated code to the action needed to apply it. The key challenge was whether a developer could maintain a continuous workflow across the editor, chat, and generated output.

In contrast, Claude Code and OpenCode contained the largest shares of visual-presentation and contrast related reports (55\% and 54\%, respectively). Many of these reports concerned terminal-oriented interactions. In terminal-based agents, AI responses, tool output, and diffs are rendered within a developer's existing terminal environment. Reports therefore focused on fixed foreground colors, hardcoded diff backgrounds, and incomplete support for terminal themes. These choices could make text difficult to read and could blur the distinction between prompts, responses, and code changes.

Cursor showed a different pattern. Readability, scaling, and interface control accounted for the largest share of its clustered reports (40\%), followed closely by visual-presentation concerns (38\%). Reports frequently concerned the chat and agent panel inside the code editor. Users described text that was too small, and settings that did not provide independent control over AI-panel appearance. Several reports also suggested that the font-size controls were difficult to find, did not reliably affect the chat surface, or required an extension~\footnote{\url{https://forum.cursor.com/t/how-can-i-increase-the-font-size-in-the-chat-window/23379/18}}. A frequent workaround was to zoom the entire application, even though this could reduce usable workspace or introduce additional layout problems. 

OpenAI Codex showed a more balanced distribution across the three categories. This pattern is consistent with its mixed interaction environment, which includes command-line and terminal interfaces, a desktop application, and an editor extension. Reports spanned screen-reader access to dynamic terminal output, contrast and visual distinction in conversations and diffs, and scaling problems in desktop and extension-based interfaces. Rather than concentrating in one surface, Codex reports reflected accessibility dependencies that moved with developers across these connected environments.

\subsubsection{Response Patterns}

As an exploratory analysis, we examined the repository level response metrics for our issue dataset. The 122 VS Code Copilot issues in our dataset came from 57 unique reporters, an average of 2.14 reports per reporter. Its five most active reporters contributed 44.3\% of the reports. In comparison, other tools averaged between 1.04 and 1.08 reports per reporter, and their five most active reporters contributed between 7.2\% and 14.0\% of reports. The Copilot pattern may reflect
sustained accessibility testing or reporting by a small group of contributors, although more information about their roles and reporting practices would be needed to interpret this pattern.

Table~\ref{tab:response_metric} summarizes response patterns in the four GitHub repositories. Approximately 87\% of Copilot and Claude Code reports were closed, compared with 84.2\% of OpenCode reports and 33.8\% of Codex reports. However, the closure processes differed. Human accounts closed most Copilot reports, whereas automated accounts closed 73.8\% of all Claude Code reports. OpenCode closures were more evenly divided between bot and human accounts, while all Codex closures were performed by human accounts. Here, a human closure refers to closure by any non-bot account and does not necessarily indicate maintainer involvement.

We also looked into the stated reason of closure. Among closed reports, 66.7\% of Codex reports, 66.0\% of Copilot reports, and 58.3\% of OpenCode reports were marked as \emph{completed}, compared with
11.3\% of Claude Code reports. A \emph{completed} closure, however, does not establish that the reported accessibility barrier was actually fixed.

Non-bot comments appeared on 77.9\% of Copilot reports, 57.7\% of Codex reports, 54.4\% of OpenCode reports, and 46.8\% of Claude Code reports. Maintainer-associated accounts participated in 66.4\% of Copilot reports, compared with 19.3\% for OpenCode, 4.2\% for Claude Code, and 2.8\% for Codex.

Human closures generally occurred sooner than automated closures. The median time to human closure was 0.3 days for Claude Code, 3.7 days for OpenCode, 4.6 days for Copilot, and 24.8 days for Codex. Median bot-closure times were 29.4 days for Claude Code, 62.2 days for Copilot, and 77.7 days for OpenCode. These longer times may reflect inactivity-based automation, although we did not analyze the automation rules used by each repository.

These measures primarily describe repository level patterns for our issue dataset, rather than the quality of accessibility remediation. Overall, the tools differed not only in the barriers visible in their reports, but also in how reporting was distributed across contributors and how the issue reports were handled.

\begin{table*}[t]
\centering
\caption{Response patterns for visual accessibility reports across GitHub-based issue tracking. Closure percentages use the total number of reports as the denominator, except Completed, which uses the number of closed reports. Human-closed refers to issues closed by a non-bot account.}
\label{tab:response_metric}
\footnotesize
\setlength{\tabcolsep}{5pt}
\begin{tabular}{lrrrrrrrr}
\toprule
 & & \multicolumn{4}{c}{Closure (\%)} & Maintainer & \multicolumn{2}{c}{Median close (days)} \\
\cmidrule(lr){3-6} \cmidrule(lr){8-9}
Tool & Issues & Closed & Bot & Human & Completed & comment (\%) & Bot & Human \\
\midrule
Claude Code      & 263 & 87.5 & 73.8 & 13.7 & 11.3 & 4.2  & 29.4 & 0.3  \\
VS Code Copilot  & 122 & 86.9 & 10.7 & 76.2 & 66.0 & 66.4 & 62.2 & 4.6  \\
OpenAI Codex     & 71  & 33.8 & 0.0  & 33.8 & 66.7 & 2.8  & --   & 24.8 \\
OpenCode         & 57  & 84.2 & 43.9 & 40.4 & 58.3 & 19.3 & 77.7 & 3.7  \\
\bottomrule
\end{tabular}
\end{table*}

\section{Discussion}

Recent user studies show that visually impaired developers can struggle to review generated changes, move between views, and follow agent activity during AI-assisted programming\cite{flores2025impact,chen2026programmers}. Our study adds a maintenance-facing perspective by showing how these broader workflow difficulties appear in public issue trackers and forums across different tools. In our dataset, editor-integrated reports more often surfaced screen-reader and semantic-access barriers. Terminal-oriented ecosystems showed larger shares of contrast and visual-presentation concerns, while reports from AI panels more often involved text size, scaling, and independent customization.

Screen-reader accessibility needs to extend beyond the host editor or terminal to the new controls and states introduced by AI assistance. AI-specific interface elements should expose a clear accessible name, role, and current state. They also need to be reachable by keyboard and provide timely feedback when the tool begins work, requires input, completes an action, or changes focus. Accessible representations of generated output should preserve the actions needed to apply, reject, or inspect that output, rather than requiring users to return to a separate visual view.

A notable finding concerns the visual distinction between human-authored and AI-generated content. Developers must distinguish instructions they wrote from the model's response. They must also identify tool status and proposed changes. When these elements use nearly identical styling, developers can lose track of an earlier request or struggle to determine what the tool has done.
Small text, low contrast, weak hierarchy, and limited zoom support can reduce usability for many developers. However, their effects may be especially consequential for low-vision and color-vision-deficient developers. Theme-aware rendering, sufficient contrast, visible hierarchy, and non-color cues can make these interactions
easier to follow. AI panels should also support independent control over text size and typography rather than relying only on application-wide zoom. 

Our exploratory response pattern analysis further shows that Copilot received more reports from a smaller group of contributors and substantially greater maintainer participation. In contrast, automated closure played a larger role in Claude Code and OpenCode. Automation may help manage large issue backlogs, but it can also make it difficult to tell whether an accessibility report received substantive review before closure. Similar overall closure rates therefore did not necessarily represent similar response processes. The concentration of Copilot reports may reflect sustained accessibility testing or advocacy, although more information about contributor roles would be needed to explain this pattern.

Accessibility barriers can also shape developers' tool choices rather than merely making a preferred workflow less convenient. One NVDA user wrote: \textit{``Personally, I currently like Cursor's AI intelligence more than that of GitHub Copilot, but I'm forced to use the latter because accessibility reasons.''}\footnote{\url{https://forum.cursor.com/t/26110}} Other reports described developers abandoning preferred IDE integrations for terminal interfaces because the terminal allowed them to enlarge text to a usable size.\footnote{\url{https://github.com/anthropics/claude-code/issues/21769}}\footnote{\url{https://github.com/openai/codex/issues/3522}} These accounts show that accessibility barriers can constrain participation in new AI-assisted development workflows.

As AI-assisted development becomes routine, these accessibility issues need to be addressed as core requirements rather than deferred usability refinements.

\section{Related Work}

\subsection{Visual Accessibility Barriers in Developer Tools}

Blind and visually impaired developers face persistent barriers across everyday programming work. Interviews and observations of blind software developers have documented difficulties with code navigation, editing, and maintaining awareness of program state when using screen readers~\cite{albusays2017interviews}. A literature review by Mountapmbeme et al.~\cite{mountapmbeme2022addressing} similarly shows that these barriers extend across the development toolchain. Storer et al.~\cite{storer2021s} found that many of the most difficult problems arise outside the code editor, including documentation, tooling, and information seeking. Even interfaces that appear primarily text-based can remain inaccessible. command-line interfaces, often assumed to be screen-reader friendly, carry their own accessibility problems~\cite{sampath2021accessibility}.

Recent work shows that AI-assisted development can both support and complicate these workflows. Flores-Saviaga et al.~\cite{flores2025impact}
studied ten visually impaired developers using GitHub Copilot to complete programming tasks. They found that AI assistance could reduce routine effort and support higher-level decision making, but frequent suggestions, context switching, and auditory interruption disrupted participants' established workflows. Through interviews with 39 blind and low-vision software professionals, Cha et al.~\cite{cha2025game} found that GenAI supported programming, debugging, documentation, and interpretation of technical visuals. However, inaccessible interfaces and the effort required to verify generated output could limit these benefits. Chen et al.~\cite{chen2026programmers} conducted a two-week study with 16 screen-reader programmers using advanced GitHub Copilot features. Their participants encountered difficulties communicating intent, reviewing generated changes, moving between views, and understanding agent status and required actions.

Together, this work shows that accessibility in AI-assisted development depends not only on access to individual controls, but also on whether developers can understand, monitor, and act on dynamic AI-generated content. Our study complements these user-centered investigations by examining how visual accessibility barriers are reported in public maintenance channels across multiple AI developer tool ecosystems.

\subsection{Mining Accessibility Related Problems}

Issue trackers, app reviews, and discussion forums often record accessibility problems as users encounter them, making them a direct source of evidence about accessibility in practice. Bi et al. characterize accessibility issues in popular GitHub projects, showing that such concerns surface widely in issue trackers and cluster around UI design, color, and navigation~\cite{bi2021first}. A separate challenge is finding these issues among large volumes of unlabeled reports. Aljedaani et al. address this by developing methods to automatically identify accessibility bug reports in open-source systems~\cite{aljedaani2022identification}. A large-scale mobile case study characterize the frequency, categories, and resolution times of accessibility bugs across the Chromium platform~\cite{aljedaani2024empirical}. App reviews carry the same signal from the end-user side. Studies have mined and classified end-user accessibility-related reviews in mobile apps~\cite{alomar2021finding,aljedaani2022automatic}. Researchers have also examined developer discussions on Stack Overflow and Twitter, surfacing recurring struggles with screen reader behavior, color contrast, and assistive technology compatibility~\cite{huq2023a11ydev, alghamdi2025understanding}. Related work has extended this mining perspective to emerging development environments. Mohammadkhani et al.~\cite{mohammadkhani2025toward} manually labeled 4,762 reviews of low-code platforms and developed a hybrid classifier to identify accessibility-related feedback. Their study shows that heavily graphical development environments can surface accessibility concerns in public user reviews. We extend this line of work by examining reports from issue trackers and forums for AI-assisted developer tools. This allows us to characterize visual accessibility barriers associated with AI-specific interaction surfaces, including chat panels, generated diffs, terminal output, and agent status interfaces.

\section{Limitations}

Our study has several limitations. Keyword-based retrieval may miss relevant reports or retain unrelated ones. The 600 reports were selected through unanimous agreement among three LLMs rather than complete manual annotation, although a stratified manual sanity check supported this conservative approach. BERTopic results depend on the selected model and parameters, and the qualitative findings are based on sampled reports rather than the full dataset. The repository metrics describe visible reporting and response activity, but issue
closure or a \emph{completed} status does not necessarily mean that an accessibility barrier was fixed. 

\section{Conclusions}

This study characterized reported visual accessibility issues across five AI developer-tool ecosystems. From a high-confidence set of 600 reports, we found that barriers extend beyond conventional screen-reader compatibility to include contrast, theme compatibility, visual differentiation, typography and scaling. The distribution of these concerns varied with the interaction surfaces emphasized by each ecosystem. Our exploratory metadata analysis further showed that GitHub-based issue tracking differed in how these reports were handled. Similar closure rates could reflect different combinations of human and automated closure, and maintainer participation varied substantially across repositories. These measures describe visible repository response patterns rather than whether the reported barriers were fixed. By highlighting both the accessibility concerns reported across AI developer tools and the ways these reports are handled, these findings can guide future research and support the design of more accessible AI developer tools.

\bibliographystyle{ACM-Reference-Format}
\bibliography{ref}

\end{document}